\documentclass[runningheads]{svmult}
\usepackage{makeidx}   
\usepackage{graphicx}  
\usepackage{subeqnar}  
\usepackage{multicol}  
\usepackage{cropmark} 
\usepackage{physprbb}  
\def\gtwid{\mathrel{\raise.3ex\hbox{$>$\kern-1.05em\lower1ex\hbox{
$\sim$}}}}
\def\ltwid{\mathrel{\raise.3ex\hbox{$<$\kern-1.05em\lower1ex\hbox{
$\sim$}}}}

\begin{document}
\title*{Why a ring of stars at $r = 20$ kpc?}
\author{Pierre Sikivie}
\institute{Department of Physics, University of Florida, 
Gainesville, FL 32611-8440, USA}

\maketitle              

\begin{abstract}
The recently discovered ring of stars near Galactocentric 
distance $r = 20$ kpc is interpreted as baryonic matter 
accreted onto the second caustic ring of dark matter in 
our galaxy.  Caustic rings of dark matter were predicted 
in the Galactic plane at radii $a_n \simeq 40~{\rm kpc}/n$ 
where $n=1,2,3 ..$.  They can reveal themselves by their 
gravitational influence on the distribution of baryonic matter. 
There is additional evidence for caustic rings of dark matter
in the Milky Way from a series of sharp rises in the Galactic 
rotation curve.  The positions of the rises are consistent at 
the 3\% level with the above law for the caustic ring radii.  
Also, a triangular feature in the IRAS map of the galactic 
plane is consistent with the imprint of the caustic ring of 
dark matter nearest to us ($n=5$) upon dust and gas in the 
Galactic plane.  These observations imply that the dark matter 
in our neighborhood is dominated by a single flow whose density 
and velocity vector are estimated.
\end{abstract}

\section{A Giant Stellar Ring in the Milky Way Plane}

Recently the discovery was announced of what appears to be 
a ring of stars, with radius of order 20 kpc, circling the 
Galaxy \cite{yan,iba,maj}.  I will not discuss the evidence 
for the alleged ring of stars, but merely assume that it 
exists.  The existence of such a ring is puzzling.  In ref. 
\cite{yan}, the ring is interpreted as the tidal stream 
from an accreted satellite galaxy.  However, in this 
interpretation it is hard to account for the fact that the 
star population is confined to a nearly circular region 
\cite{iba}.  Also, it would be an accident that the tidal 
stream lies in the Galactic plane.  Ibata et al. \cite{iba} 
propose instead that the ring of stars is a perturbation 
of the Galactic disk population.  They do not however explain 
what causes the perturbation.  I would like to propose that 
the perturbation that causes the ring of stars at $r = 20$ kpc
is the gravitational field of a caustic ring of dark matter.

Several years ago caustic rings of dark matter were predicted 
in the Galactic plane at radii $a_n \simeq 40~{\rm kpc}/n$ where 
$n = 1,2,3, ..$  \cite{cr}.  The ring of stars at $r = 20$ kpc 
may be baryonic matter accreted onto the $n=2$ caustic ring of 
dark matter.  This interpretation would account for both its 
radius and its location in the Galactic plane.

The existence of dark matter caustics in the halos of galaxies is 
a corollary of the observation \cite{ips} that cold dark matter 
particles lie on a three-dimensional sheet in six-dimensional 
phase-space.  There are compelling reasons to believe that the 
dark matter of the universe is constituted in large part of 
non-baryonic collisionless particles with very small primordial 
velocity dispersion, such as axions and/or weakly interacting 
massive particles (WIMPs).  Generically, such particles are 
called cold dark matter (CDM).  The primordial velocity dispersion 
of the leading cold dark matter candidates is extremely small, of 
order $10^{-12}c$ for WIMPs and $3\cdot 10^{-17}c$ (at most) for 
axions.  This means that, before the onset of structure formation, 
all the particles at a given location $\vec{x}$ have the same 
velocity $\vec{v}(\vec{x})$, i.e. the particles lie on a 3-dim. 
sheet in 6-dim. phase space.  The thickness of the sheet is 
the velocity dispersion.  Because the number of particles is 
huge (of order $10^{84}$ axions and/or $10^{68}$ WIMPs per 
galactic halo), the sheet is effectively continuous.  It cannot 
break and hence its evolution is constrained by topology.  

\section{The Phase-Space Structure of Galactic Halos}

Where a galaxy forms, the sheet wraps up in phase-space, turning clockwise
in any two dimensional cut $(x, \dot{x})$ of that space.  $x$ is the
physical space coordinate in an arbitrary direction and $\dot{x}$ its
associated velocity.  The outcome of this process is a discrete set of
flows at any physical point in a galactic halo \cite{ips}.  Two flows
are associated with particles falling through the galaxy for the first
time ($n=1$), two other flows are associated with particles falling
through the galaxy for the second time ($n=2$), and so on.  Scattering
in the gravitational wells of inhomogeneities in the galaxy (e.g.
molecular clouds and globular clusters) are ineffective in thermalizing
the flows with low values of $n$.  Recently, Stiff and Widrow \cite{widr}
have put these discrete flows, also called 'velocity peaks', into evidence
in $N$-body simulations using a new technique which increases the
resolution of the simulations in the relevant regions of phase-space.

A commonly raised objection to the above picture is that before
the dark matter falls onto a large galaxy, such as our own, it
has already clustered on smaller scales, making dwarf halos and
other types of clumps, in a process called "hierarchical
clustering".  However, this is not a valid objection.  The
effect of hierarchical clustering is only to produce an
effective velocity dispersion for the infalling dark matter,
i.e. a thickening of the phase space sheet.  This effective
velocity dispersion is at most equal to the velocity
dispersion, of order 10 km/s, of dwarf halos and on average
should be much less than that.  Because the effective velocity
dispersion of the infalling dark matter is much less than the
300 km/s velocity dispersion of the Galaxy as a whole, the
phase space sheet folds in qualitatively the same way as in
the zero velocity dispersion case.  The flows and caustics
remain.  

Caustics appear wherever the projection of the phase-space sheet onto
physical space has a fold \cite{cr,sing,lens,Tre}.  Generically, caustics
are surfaces in physical space.  On one side of the caustic surface
there are two more flows than on the other.  At the surface, the dark
matter density is very large.  It diverges there in the limit of zero
velocity dispersion.  There are two types of caustics in the halos of 
galaxies, inner and outer.  The outer caustics are simple fold $(A_2)$
catastrophes located on topological spheres surrounding the galaxy.  
They occur near where a given outflow reaches its furthest distance 
from the galactic center before falling back in.  The inner caustics 
are rings \cite{cr}.  They occur near where the particles with 
the most angular momentum in a given inflow reach their distance of
closest approach to the galactic center before going back out.  A 
caustic ring is a closed tube whose cross-section is an {\it elliptic 
umbilic} ($D_{-4}$) catastrophe \cite{sing}.  The existence of these 
caustics and their topological properties are independent of any 
assumptions of symmetry.

Primordial peculiar velocities are expected to be the same for baryonic
and dark matter particles because they are caused by gravitational forces.
Later the velocities of baryons and CDM differ because baryons collide
with each other whereas CDM is collisionless. However, because angular
momentum is conserved, the net angular momenta of the dark matter and
baryonic components of a galaxy are aligned.  Since the caustic rings
are located near where the particles with the most angular momentum in
a given infall are at their closest approach to the galactic center,
they lie close to the galactic plane.

\begin{figure}[b]
\begin{center}
\vskip .5cm
\includegraphics[width=.6\textwidth]{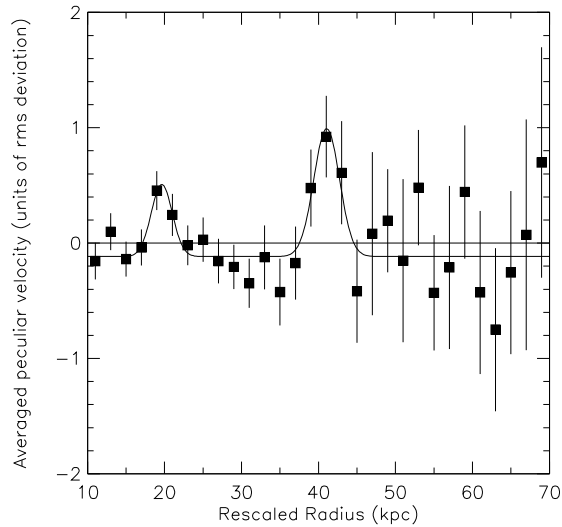}
\end{center}
\caption[]{Composite rotation curve constructed in ref. \cite{kinn}.
It combines data on 32 exterior galaxies to test the hypothesis that 
the caustic ring radii are given by Eq. (\ref{crr}).}
\label{comprot}
\end{figure}

A specific proposal was made for the radii $a_n$ of caustic rings
\cite{cr}:
\begin{equation}
\{a_n: n=1,2, ...\} \simeq (39,~19.5,~13,~10,~8,...) {\rm kpc}
\times \left({j_{\rm max}\over 0.25}\right) 
\left({v_{\rm rot} \over 220 {{\rm km} \over {\rm s}}} \right)
\label{crr}
\end{equation}
where $v_{\rm rot}$ is the rotation velocity of the galaxy and 
$j_{\rm max}$ is a parameter with a specific value for each halo.  
For large $n$, $a_n \propto 1/n$.  Eq. \ref{crr} is predicted by
the self-similar infall model \cite{ss,sty} of galactic halo formation.
$j_{\rm max}$ is then the maximum of the dimensionless angular momentum
$j$-distribution \cite{sty}.  The self-similar model depends upon a
parameter $\epsilon$ \cite{ss}.  In CDM theories of large scale structure
formation, $\epsilon$ is expected to be in the range 0.2 to 0.35
\cite{sty}. Eq. \ref{crr} is for $\epsilon = 0.3$.  However, in the 
range $0.2 < \epsilon < 0.35$, the ratios $a_n/a_1$ are almost
independent of $\epsilon$.  When $j_{\rm max}$ values are quoted
below, $\epsilon = 0.3$ and $h = 0.7$ will be assumed.

It was pointed out in ref. \cite{sty} that including angular momentum
in the self-similar infall model results in a depletion of the inner
halo and hence an effective core radius.  The average amount of angular
momentum of the Milky Way halo was estimated \cite{sty} by requiring that
approximately half of the rotation velocity squared at our location is due
to dark matter, the other half being due to ordinary matter.  This yields
$\bar{j} \sim 0.2$ where $\bar{j}$ is the average of the $j$-distribution
for our halo.  $\bar{j}$ and $j_{\rm max}$ are related if some assumption
is made about the shape of the $j$-distribution.  For example, if the
$j$-distribution is taken to be that of a rigidly rotating sphere,
one has $j_{\rm max} = {4 \over \pi} \bar{j}$. Hence
$j_{\rm max} \sim 0.25$ for our halo.

Since caustic rings lie close to the galactic plane, they cause
bumps in the rotation curve, at the locations of the rings.  In
ref. \cite{kinn} a set of 32 extended well-measured rotation curves
was analyzed and statistical evidence was found for the $n=1$ and
$n=2$ caustic rings, distributed according to Eq. \ref{crr}.  In 
this analysis, each of the 32 individual galactic rotation curves 
was rescaled according to 
\begin{equation}
r \rightarrow 
\tilde{r} = r \left({220~{\rm km/s} \over v_{\rm rot}}\right)
\label{resc}
\end{equation}
where $v_{\rm rot}$ is the measured rotation velocity.  To isolate 
the outer halo-dominated portion of the rotation curves, all data 
with rescaled radii $\tilde{r} < 10$ kpc were removed.  Each rotation 
curve was then fitted to a line or a quadratic polynomial.  The residual
deviations were normalized to the rms deviation in each fit and then 
binned together.  The result is the composite rotation shown in 
Fig. \ref{comprot} for the case where the individual rotation curves 
were fitted to quadratic polynomials. The composite rotation curve has 
two peaks, near 20 kpc and 40 kpc, with statistical significance of 
$3\sigma$ and $2.6\sigma$ respectively.  It implies that the 
$j_{\rm max}$ distribution is peaked near 0.27.  The rotation 
curve of NGC3198, one of the best measured, by itself shows 
three faint bumps which are consistent with Eq. \ref{crr} and 
$j_{\rm max} = 0.28$~\cite{cr}.   Also our earlier estimate of
$j_{\rm max}$ for the Milky Way halo is close to the peak value 
of 0.27.

\section{Evidence for Ring Caustics in the Milky Way}

Eq. \ref{crr} with $j_{\rm max} = 0.25$ implies that our halo has
caustic rings with radii near 40 kpc/$n$, where $n$ is an integer.
Here we point to evidence in support of this extraordinary claim.  

It was already mentioned that the recently discovered ring of 
stars at $r = 20$ kpc may be interpreted as baryonic matter 
accreted onto the $n=2$ caustic ring of dark matter.  This 
interpretation accounts for the 20 kpc radius of the ring as 
well as for its location in the galactic plane.  The spatial 
coincidence of the 20 kpc ring of stars with the predicted 
$n=2$ caustic ring may, of course, be fortuitous.  So it is 
natural to ask whether there is similar evidence for any of
the other caustic rings.  The answer is yes for the $n=3$ ring.
Binney and Dehnen studied \cite{bin} the outer rotation curve of
the Milky Way and concluded that its anomalous behavior can be
explained if most of the tracers of the rotation are concentrated
in a ring of radius $1.6~r_\odot$ where $r_\odot$ is our distance to
the galactic center.  Throughout this paper we use the standard value
$r_\odot = 8.5$ kpc.  That value is assumed in Eq. \ref{crr}, and
also in refs. \cite{yan,iba}.  The Binney and Dehnen ring is therefore
at 13.6 kpc, which is very close to the predicted radius of the $n=3$
caustic ring.  Moreover, there is independent evidence for the
existence of the Binney and Dehnen ring.

\begin{figure}[b]
\begin{center}
\vskip 3.5cm
\includegraphics[width=.466\textwidth]{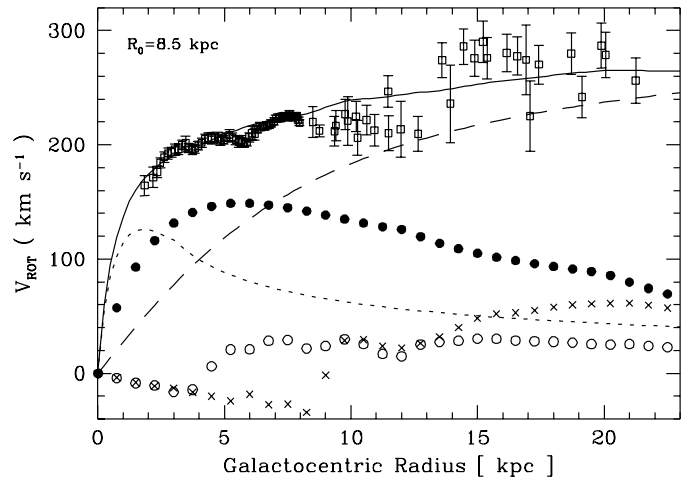}
\end{center}
\vskip -3.5cm
\caption[]{Milky Way rotation curve from ref. \cite{olli}.  The 
different lines represent the contributions from the bulge (dotted), 
the stellar disk (filled circles), the HI layer (crosses), the H$_2$ 
layer (circles), and from a smooth dark halo (dashed).  The full line
represents the sum of the contributions.  Reprinted by permission of 
the authors and Blackwell Publishing Ltd.}
\label{outrot}
\end{figure}

Olling and Merrifield have recently published \cite{olli} a rotation
curve for the Milky Way.  It is reproduced in Fig. \ref{outrot}.
It shows a significant rise between 12.7 and 13.7 kpc.  The increase
in rotation velocity is 27\%, from 220 to 280 km/s.  A ring of matter
in the Galactic plane produces a rise the rotation curve.  The rise
expected from the $n=3$ caustic ring of dark matter, by itself, is
only of order 3\%.  However, the effect of a caustic ring of dark
matter is amplified by the ordinary matter (stars, gas, dust ..)
which it attracts gravitationally.  The amplification would have to
be by a factor of order nine in the case of the $n=3$ ring.  One may
think, at first, that such as a large amplification is implausible
because the back reaction of the ring of ordinary matter upon the
caustic ring of dark matter would determine the position of the
latter, instead of the latter determining the position of the former.
However this is not so because the dark matter caustic is not an
overdensity of particles which are at rest with respect to the
caustic ring.  The particles which at a given time make up the
caustic ring are moving with great speed, of order 360 km/s for
$n=3$, and are continually replaced by new particles.  As a result,
the position of the caustic ring is insensitive to the gravitational
field of the matter it attracts.

The existence of rings of ordinary matter precisely where the $n=2$
and $n=3$ rings of dark matter are predicted may yet be fortuitous.
Fig. 1 does not show a significant rise near the predicted location
(10 kpc) of the $n=4$ caustic ring.  Note however that the error
bars in Fig. 1 are very large for $r > r_\odot$.  The rise near
10 kpc, if indeed there is one, may be too small to show up in
the data.  On the other hand, the inner ($r < r_\odot$) part of
the rotation curve is far better measured, and we may go look for
rises there.

\begin{figure}[b]
\vskip 1.8cm
\begin{center}
\includegraphics[width=.5\textwidth]{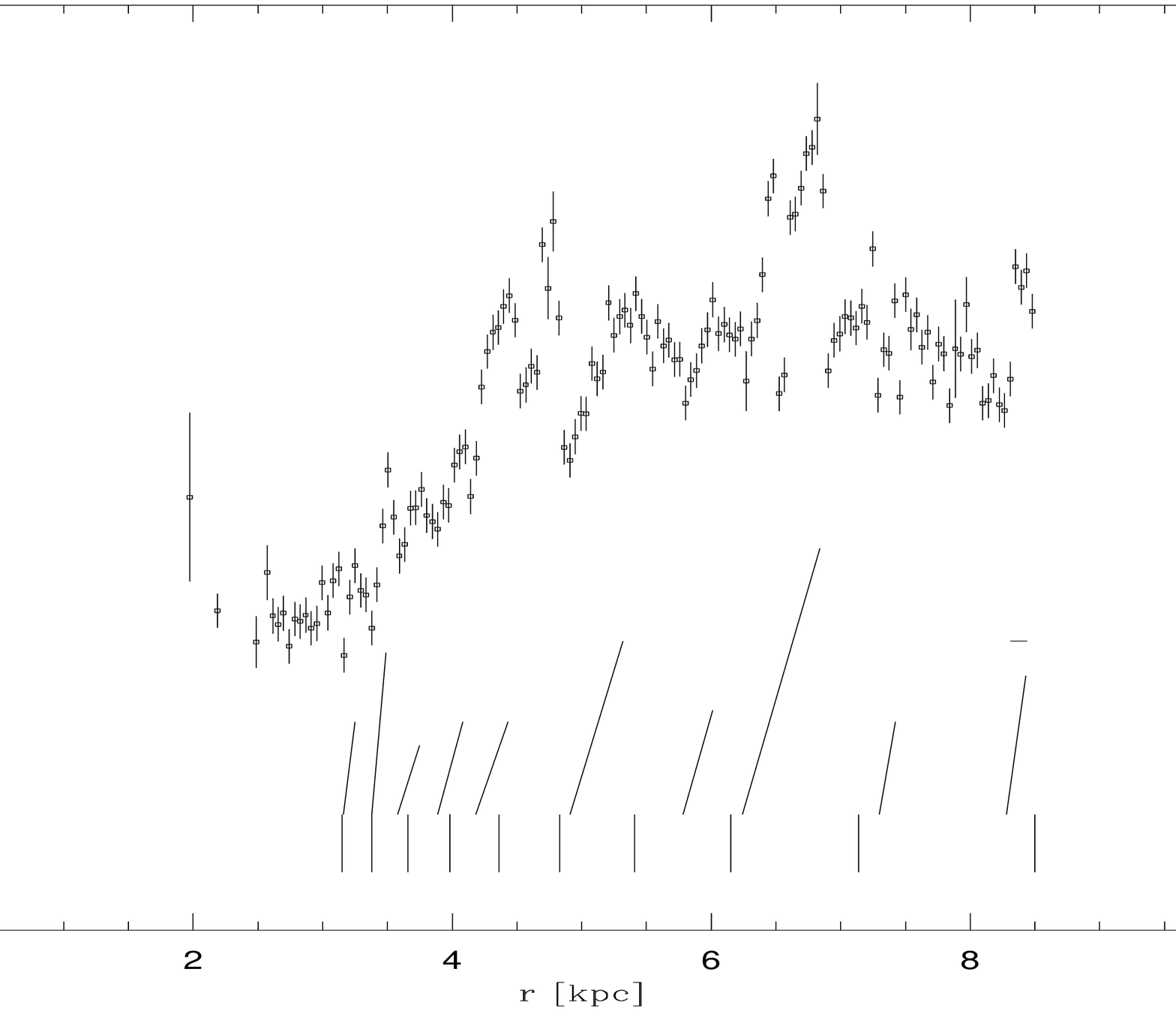}
\end{center}
\vskip .2cm
\caption[]{North Galactic rotation curve from ref. \cite{clem}.  The
locations of the rises listed in the first three columns of Table I are
indicated by line segments parallel to the rises but shifted downwards.  
The caustic ring radii for the fit described in the text are shown as
vertical line segments.  The position of the triangular feature in
the IRAS map of the galactic plane near $80^\circ$ longitude is shown
by the short horizontal line segment.  It coincides with a rise in the
rotation curve.}
\label{inrot}
\end{figure}

Galactic rotation curves are obtained from HI and CO surveys of
the Galactic plane.  A list of surveys performed to date is given
in ref. \cite{merr}.  Everything else being equal, CO surveys
have far better angular resolution than HI surveys because their
wavelength is nearly two orders of magnitude smaller (0.26 cm vs.
21 cm).  The most detailed inner Galactic rotation curve appears
to be that obtained \cite{clem} from the Massachusetts-Stony Brook
North Galactic Plane CO survey \cite{CO}.  It is reproduced in
Fig. \ref{inrot}.  It shows highly significant rises between 3
and 8.5 kpc.  Eq. \ref{crr} predicts ten caustic rings between
3 and 8.5 kpc.  Allowing for ambiguities in identifying rises,
the number of rises in the rotation curve between 3 and 8.5 kpc
is in fact approximately ten.  Below 3 kpc the predicted rises
are so closely spaced that they are unlikely to be resolved in
the data.

\begin{table}
\caption{Radii at which rises in the Milky Way rotation curve of
Fig. \ref{inrot} start ($r_1$) and end ($r_2$), the corresponding 
increases in velocity $\Delta v$, the caustic ring radii $a_n$ of the
self-similar infall model for the fit described in the text, and typical
velocity increases $\bar{\Delta} v_n$ predicted by the model without
amplification due to accretion of ordinary matter onto the caustic rings.}
\vspace{0.5cm}
\begin{center}
\renewcommand{\arraystretch}{1.4}
\setlength\tabcolsep{15pt}
\begin{tabular}{ccccc}
$r_1$  &$r_2$&$\Delta v$ &    $a_n$        &$\bar{\Delta}v_n$\\
(kpc)  & (kpc) & (km/s) &     (kpc)        & (km/s)      \\
       &       &        &   n = 1 ..14     &  n = 1 .. 14  \\
\hline
       &       &       &      41.2         & 26.5  \\
       &       &       &      20.5         & 10.6  \\
       &       &       &      13.9         &  6.8  \\
       &       &       &      10.5         &  5.0  \\
8.28   & 8.43  &   12  &       8.50        &  3.9  \\
(7.30) &(7.42) &   (8) &       7.14        &  3.2  \\
6.24   & 6.84  &   23  &       6.15        &  2.6  \\
5.78   & 6.01  &    9  &       5.41        &  2.3  \\
4.91   & 5.32  &   15  &       4.83        &  2.0  \\
4.18   & 4.43  &    8  &       4.36        &  1.7  \\
3.89   & 4.08  &    8  &       3.98        &  1.5  \\
3.58   & 3.75  &    6  &       3.66        &  1.3  \\
3.38   & 3.49  &   14  &       3.38        &  1.2  \\
3.16   & 3.25  &    8  &       3.15        &  1.1  \\
\hline
\end{tabular}
\end{center}
\label{tbl1}
\end{table}

Table I lists ten rises, identified by the radius $r_1$ where they
start, the radius $r_2$ where they end, and the increase $\Delta v$
in rotation velocity.  The rises are marked as slanted line segments
in Fig. \ref{inrot}.  The evidence for the rise between 7.30 and
7.42 kpc is relatively weak, so the corresponding numbers are
in parenthesis in the Table.

The effect of a caustic ring in the plane of a galaxy upon its
rotation curve was analyzed in ref. \cite{sing}.  The caustic
ring produces a rise in the rotation curve which starts at
$r_1 = a_n$, where $a_n$ is the caustic ring radius, and which
ends a $r_2 = a_n + p_n$, where $p_n$ is the caustic ring
width.  The ring widths depend in a complicated way on the
velocity distribution of the infalling dark matter at last
turnaround \cite{sing} and are not predicted by the model.
They also need not be constant along the ring.

In the past, rises (or bumps) in galactic rotation curves have
been interpreted as due to the presence of spiral arms \cite{spam}.
Spiral arms may in fact cause some of the rises in rotation curves.
This does not, however, exclude the possibility of other valid
explanations.  Two properties of the high resolution rotation
curve of Fig. \ref{inrot} favor the interpretation that its
rises are caused by caustic rings of dark matter.  First, there
are of order ten rises in the range of radii covered (3 to 8.5 kpc).
This agrees qualitatively with the predicted number of caustic rings,
whereas only three spiral arms are known in that range: Scutum,
Sagittarius and Local. Second, the rises are sharp transitions
in the rotation curve, both where they start $(r_1)$ and where
they end $(r_2)$.  Sharp transitions are consistent with caustic
rings because the latter have divergent density at $r_1 = a$ and
$r_2 = a + p$ in the limit of vanishing velocity dispersion.
Finally, there are bumps and rises in rotation curves measured
at galactocentric distances much larger than the disk radius,
where no spiral arms are seen.  In particular, the features
found in the composite rotation curve of Fig. \ref{comprot}
are at distances 20 kpc and 40 kpc when scaled to our own galaxy.

The self-similar infall model prediction for the caustic ring radii,
Eq. \ref{crr}, was fitted to the eight rises between 3 and 7 kpc by
minimizing $rmsd \equiv [{1 \over 8} {\displaystyle \sum_{n=7}^{14}}
( 1 - {a_n \over r_{1 n}})^2]^{1 \over 2}$ with respect to $j_{\rm max}$,
for $\epsilon = 0.30~$.  The fit yields $j_{\rm max} = 0.263$ and
$rmsd = 3.1\%~$. The fourth column of Table I shows the corresponding
caustic ring radii $a_n$.  In Fig. \ref{inrot} the latter are indicated
by short vertical line segments.

The velocity increase due to a caustic ring is given by
\begin{equation}
\Delta v_n = v_{\rm rot} f_n {\Delta I(\zeta_n)
\over \cos\delta_n(0) + \phi_n^\prime(0) \sin\delta_n(0)}\ .
\label{dv}
\end{equation}
The $f_n$, defined in ref. \cite{cr}, are predicted by the self-similar
infall model, but $\Delta I(\zeta_n), \delta_n(0)$ and $\phi_n^\prime(0)$,
defined in ref. \cite{sing}, are not.  Like the $p_n$, the latter
parameters depend in a complicated way on the velocity distribution
of the dark matter at last turnaround.  On the basis of the discussion
in ref. \cite{sing}, the ratio on the RHS of Eq. \ref{dv} is expected
to be of order one, but to vary from one caustic ring to the next.  The
size of these fluctuations is easily a factor two, up or down.  The
fifth column of Table I shows $\Delta v_n$ with the fluctuating ratio
set equal to one, i.e. $\bar{\Delta} v_n \equiv f_n v_{\rm rot}$.

For the reasons just stated, the fact that the observed $\Delta v$
fluctuate by a factor of order 2 from one rise to the next is
consistent with the interpretation that the rises are due to
caustic rings.  However the observed $\Delta v$ (column 3) are
typically a factor 5 larger than the velocity increases expected
from the caustic rings acting alone (column 5).  This is similar
to what we found for the $n=3$ caustic ring, and suggests
that the effects of the $n = 5 ... 14$ caustic rings are also
amplified by baryonic matter they have accreted.  I'll argue
that the gas in the disk has sufficiently high density and low
velocity dispersion to produce such large amplification factors.
I'll also give observational evidence in support of the hypothesis.

The equilibrium distribution of gas is:
\begin{equation}
d_{\rm gas} (\vec{r}) = d_{\rm gas}(\vec{r}_0)
\exp[- {3 \over <v^2_{\rm gas}>}(\phi(\vec{r}) - \phi(\vec{r}_0))]\ ,
\label{gas}
\end{equation}
where $d$ is density and $\phi$ gravitational potential.  In the
solar neighborhood, $d_{\rm gas} \simeq
3\cdot 10^{-24} {{\rm gr} \over {\rm cm}^3}$ \cite{BT}, which is
comparable to the density of dark matter inside the tubes of
caustic rings near us.  From the scale height of the gas \cite{BT}
and the assumption that it is in equilibrium with itself and the
other disk components, I estimate
$\langle v_{\rm gas}^2 \rangle ^{1 \over 2} \simeq 8$ km/s.  The
variation in the gravitational potential due to a caustic ring over
the size of the tube is of order $\Delta \phi_{\rm CR} \simeq
2 f v_{\rm rot}^2 {p/a} \simeq (5 {{\rm km} \over {\rm s}})^2$.
Because ${3 \over <v_{\rm gas}^2>} \Delta \phi_{\rm CR}$ is of order
one, the caustic rings have a large effect on the distribution of
gas in the disk.  The accreted gas amplifies and can dominate the
effect of the caustic rings on the rotation curve.  To check
whether this hypothesis is consistent with the shape of the
rises would require detailed modeling, as well as detailed
knowledge on how the rotation curve is measured.  However,
there is observational evidence in support of the hypothesis.

The accreted gas may reveal the location of caustic rings in maps of the
sky.  Looking tangentially to a ring caustic from a vantage point in the
plane of the ring, one may recognize the tricusp \cite{sing} shape of the
$D_{-4}$ catastrophe.  I searched for such features.  The IRAS map of the
galactic disk in the direction of galactic coordinates $(l,b) = (80^\circ,
0^\circ)$ shows a triangular shape which is strikingly reminiscent of the
cross-section of a ring caustic.  The relevant IRAS maps are posted at
http://www.phys.ufl.edu/$\sim$sikivie/triangle/ .  They were downloaded
from the Skyview Virtual Observatory (http://skyview.gsfc.nasa.gov/).
The vertices of the triangle are at $(l,b) = (83.5^\circ, 0.4^\circ),
(77.8^\circ, 3.4^\circ)$ and $(77.8^\circ, -2.6^\circ)$.   The shape
is correctly oriented with respect to the galactic plane and the galactic
center.  To an extraordinary degree of accuracy it is an isosceles
triangle with axis of symmetry parallel to the galactic plane, as is
expected for a caustic ring whose transverse dimensions are small compared
to its radius.  Moreover its position is consistent with the position of a
rise in the rotation curve, the one between 8.28 and 8.43 kpc ($n=5$).
The caustic ring radius implied by the image is 8.31 kpc and its
dimensions are $p \sim 130$ pc and $q \sim 200$ pc, in the
directions parallel and perpendicular to the galactic plane
respectively.  It therefore predicts a rise which starts at 8.31 kpc
and ends at 8.44 kpc, just where a rise is observed.  The probability
that the coincidence in position of the triangular shape with a rise in
the rotation curve is fortuitous is less than $10^{-3}$.

In principle, the feature at $(80^\circ, 0^\circ)$ should be matched
by another in the opposite tangent direction to the nearby ring caustic,
at approximately $(-80^\circ, 0^\circ)$.  Although there is a plausible
feature there, it is much less compelling than the one in the
$(+80^\circ, 0^\circ)$ direction.  There are several reasons why
it may not appear as strongly.  One is that the $(+80^\circ, 0^\circ)$
feature is in the middle of the Local spiral arm, whose stellar
activity enhances the local gas and dust emissivity, whereas the
$(-80^\circ, 0^\circ)$ feature is not so favorably located.  Another
is that the ring caustic in the $(+80^\circ, 0^\circ)$ direction has
unusually small dimensions.  This may make it more visible by increasing
its contrast with the background.  In the $(-80^\circ,0^\circ)$ direction,
the nearby ring caustic may have larger transverse dimensions.

\section{The Big Flow}

Our proximity to a caustic ring means that the corresponding flows, i.e.
the flows in which the caustic occurs, contribute very importantly to the
local dark matter density.  Using the results of refs. \cite{cr,sing,sty},
we can estimate their densities and velocity vectors.  Let us assume, for
illustrative purposes, that we are in the plane of the nearby caustic
and that its outward cusp is 55 pc away from us, i.e. $a_5 + p_5 =
8.445$ kpc.  The densities and velocity vectors on Earth of the $n=5$
flows are then:
\begin{equation}
d^+ = 1.7~10^{-24}~{{\rm gr} \over {\rm cm}^3}~,~
d^- = 1.5~10^{-25}~{{\rm gr} \over {\rm cm}^3}~,~
\vec{v}^\pm = (470~\hat{\phi} \pm ~100~\hat{r})~
{{\rm km} \over {\rm s}},
\label{lc}
\end{equation}
where $\hat{r}, \hat{\phi}$ and $\hat{z}$ are the local unit vectors
in galactocentric cylindrical coordinates.  $\hat{\phi}$ is in the
direction of galactic rotation. The velocities are given in the
(non-rotating) rest frame of the Galaxy.  Because of an ambiguity,
it is not presently possible to say whether $d^\pm$ are the
densities of the flows with velocity $\vec{v}^\pm$ or $\vec{v}^\mp$.
The large size of $d^+$ is due to our proximity to the outward cusp
of the nearby caustic.  Its exact value is sensitive to our distance
to the cusp.  We do not know that distance well enough to estimate
$d^+$ with accuracy.  However we can say that $d^+$ is very large,
of order the value given in Eq. \ref{lc}, perhaps even larger.  If we
are inside the tube of the nearby caustic, there are two additional
flows on Earth, aside from those given in Eq. \ref{lc}.  A list of
local densities and velocity vectors for the $n \neq 5$ flows can
be found in ref. \cite{bux}. An updated list is in preparation.

Eq. \ref{lc} has dramatic implications for dark matter searches.
Previous estimates of the local dark matter density, based on
isothermal halo profiles, range from 5 to
7.5~$10^{-25}~{{\rm gr} \over {\rm cm}^3}$.  The present analysis
implies that a single flow ($d^+$) has of order three times (or
more) that much local density.

The sharpness of the rises in the rotation curve and of the triangular
feature in the IRAS map implies an upper limit on the velocity dispersion
$\delta v_{\rm DM}$ of the infalling dark matter.  Caustic ring
singularities are spread over a distance of order
$\delta a \simeq {R~\delta v_{\rm DM} \over v}$ where $v$ is the 
velocity of the particles in the caustic, $\delta v_{\rm DM}$ is 
their velocity dispersion, and $R$ is their turnaround radius.  The
sharpness of the IRAS feature implies that its edges are spread over
$\delta a \ltwid 20$ pc.  Assuming that the feature is due
to the $n=5$ ring caustic, $R \simeq$ 180 kpc and $v \simeq 480$ km/s.
Therefore $\delta v_{\rm DM} \ltwid 53$ m/s.  This bound is strongly
at odds with the often quoted claim that structure formation on small
scales causes all, or nearly all, late infalling dark matter to be in
dwarf galaxy type clumps, with velocity dispersion of order 10 km/s.

The caustic ring model may explain the puzzling persistence of galactic
disk warps \cite{war}.  These may be due to outer caustic rings lying
somewhat outside the galactic plane and attracting visible matter.
The resulting disk warps would not damp away, as is the case in
more conventional explanations of the origin of the warps, but
would persist on cosmological time scales.

The caustic ring model, and more specifically the prediction Eq.~\ref{lc}
of the locally dominant flow associated with the nearby ring, has
important consequences for axion dark matter searches \cite{adm},
the annual modulation \cite{bux,ann,stiff,SW} and signal anisotropy
\cite{anis,stiff} in WIMP searches, the search for $\gamma$-rays
from dark matter annihilation \cite{gam}, and the search for
gravitational lensing by dark matter caustics \cite{lens,mou}.
The model makes predictions for each of these approaches to the
dark matter problem.

I thank the Aspen Center for Physics for its hospitality while 
this paper was written.  This work is supported in part by 
U.S. DOE grant DEFG05-86ER-40272.

\end{document}